\documentclass{iau}
\usepackage{graphicx}

\title[All Sky Young Association (ASYA)] 
{The All Sky Young Association (ASYA):\\ a new young association}

\author[Torres, Quast \& Montes]   
{
C. A. O. Torres$^{1}$, 
G. R. Quast$^{1}$,
\and D.~Montes$^{2}$, 
 }

\affiliation{$^1$ Laborat\'orio Nacional de Astrof\'{\i}sica/ MCT, Rua Estados Unidos 154, 37504-364, Itajub\'a, MG, Brazil email: {\tt beto@lna.br}\\
$^2$Dpto. Astrof\'{\i}sica, Facultad de CC. F\'{\i}sicas, Universidad Complutense de Madrid, E-28040 Madrid, Spain\\email: {\tt dmontes@ucm.es}
}

\pubyear{2015}
\volume{314}  
\pagerange{119--126}
\jname{Young Stars \& Planets Near the Sun}
\editors{J. H. Kastner, B. Stelzer, \& S. A. Metchev, eds.}
\begin{document}

\maketitle

\begin{abstract}
To analyze the SACY (Search for Associations Containing Young stars) survey we developed a method to find young associations and to define their high probability members. These bona fide members enable to obtain the kinematical and the physical properties of each association in a proper way. Recently we noted a concentration in the $UV$ plane and we found a new association we are calling ASYA (All Sky Young Association) for its overall distribution in the sky with a total of 38 bonafide members and an estimated  age of 110 Myr, the oldest young association found in the SACY survey.  We present here its kinematical, space and Li distributions and its HR diagram. 
\end{abstract}

\firstsection 
\section{Introduction}

The SACY (Search for Associations Containing Young Stars) survey was a spectroscopic effort to find southern nearby associations using 
as targets the possible Tycho--2/{\sc Hipparcos} stars  counterparts of the ROSAT All-Sky Bright Sources Catalogue (\cite{Torres06}; \cite[2008]{Torres08}). 
To analyze the SACY survey we developed a method to find young associations and to define their high probability members. These bonafide members enable to obtain the kinematical and the physical properties of each association in a proper way.

\section{ASYA (All Sky Young Association)}

\begin{figure}[h]
 \vspace*{-0.2 cm}
\begin{center}
\centerline{
\includegraphics[width=5.0in]{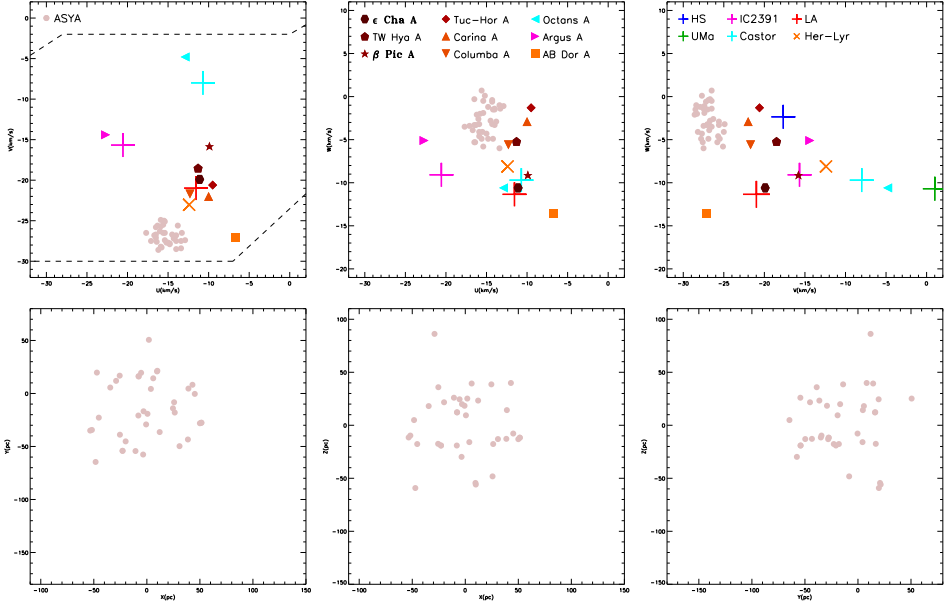}
}
 \vspace*{-0.2 cm}
 \caption{
\textit{Upper panel}:  $UVW$ space for ASYA showing the well defined kinematical clustering. 
The position of the classical moving groups 
(Hyades Supercluster (HS), Ursa Major (UMa), IC2391, Castor,  Local Association (LA) and Hercules-Lyra, 
\cite{mon01} and \cite{mon10}) 
and the young associations of SACY ($\epsilon$ Cha, TW Hya, $\beta$ Pic, Tuc-Hor, Carina, Columba, Octans, Argus and AB Dor) 
are also presented for comparison. Note that ASYA is clearly separated from Her-Lyr and the other younger associations.
\textit{Lower panel}:  $XYZ$ space for ASYA.
}
   \label{fig:uvw_xyz}
\end{center}
\end{figure}

\begin{figure}[h]
 \vspace*{-0.4 cm}
\begin{center}
\centerline{
\includegraphics[width=5.0in]{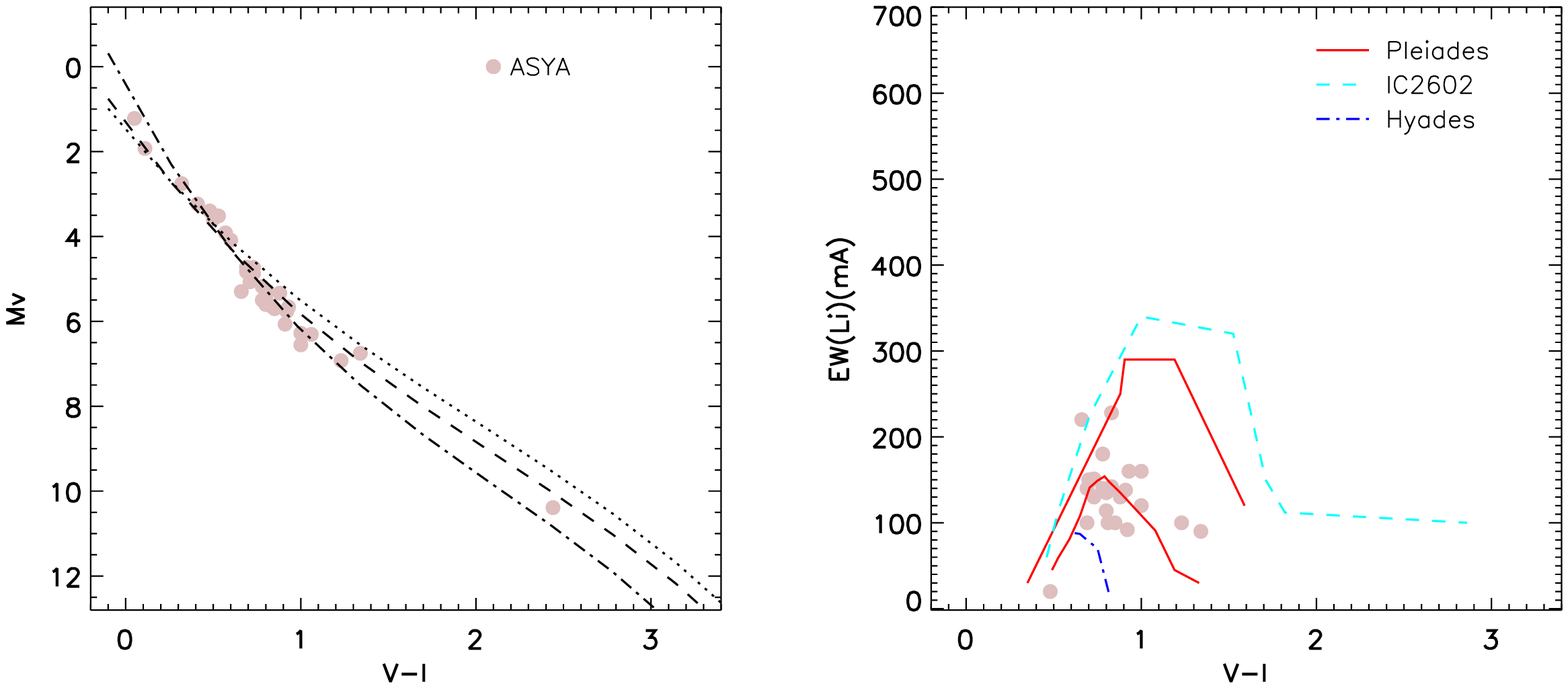}
}
 \vspace*{-0.2 cm}
 \caption{
 \textit{Left panel}: The HR diagram for the proposed members of ASYA. The over-plotted curves are our ad-hoc isochrones. The upper curve is estimated as about 20 Myr and the lower curve 110 Myr. 
\textit{Right panel}: Lithium distribution ($EW$(Li) vs $V-I$) of ASYA compared with the Li distribution of IC~2602, Pleiades and Hyades open clusters. 
}
   \label{fig:HR_Li}
\end{center}
\end{figure}

Recently we noted a concentration in the $UV$ plane and we found a new association we are calling ASYA (All Sky Young Association) for its overall distribution in the sky. We also search {\sc Hipparcos} catalogue for other possible members and we found a total of 38 bonafide members (including those from {\sc Hipparcos} and SACY). Although kinematically 
($U=-15.2,
V=-26.9,
W=-2.8$ km/s) 
near the Her-Lyr moving group, ASYA is definitively distinct from it and younger - we estimated an age of 110 Myr, the oldest of the young associations found in the SACY survey. A weak expansion in the X direction is present $U=<U> +k(X-<X>)$ and the solution was obtained with a $k=0.02$.
We present its kinematical ($U V W$) and space ($XYZ$)  distributions in Fig.~\ref{fig:uvw_xyz}  
and its Li distribution ($EW$(Li) vs $V-I$) and its HR diagram ($M_{V}$ vs. $V-I$) in Fig.~\ref{fig:HR_Li}.

\begin{acknowledgements}
This work was supported by the Univ. Complutense de Madrid (UCM) and the  Ministry of Economy and Competitiveness (MINECO) under grant AYA2011-30147-C03-02.
\end{acknowledgements}


%


\end{document}